\documentclass[aps,a4paper,superscriptaddress,twocolumn]{revtex4}
\usepackage{tensor}
\usepackage{graphicx}
\usepackage{amsmath}
\usepackage{amssymb}
\usepackage{enumerate}
\usepackage{subfigure}
\usepackage{tabularx}
\usepackage[colorlinks=true, pdfstartview=FitV, linkcolor=blue, citecolor=red, urlcolor=black, breaklinks=true]{hyperref}
\newcommand{\be}{\begin{equation}}
\newcommand{\ee}{\end{equation}}
\newcommand{\ben}{\begin{eqnarray}}
\newcommand{\een}{\end{eqnarray}}
\newcommand{\bes}{\begin{subequations}}
\newcommand{\ees}{\end{subequations}}
\def\bal#1\eal{\begin{align}#1\end{align}}

\newcommand{\sech}{{\rm sech}}
\newcommand{\LL}{{\mathcal L}}

\newcommand{\veps}{\varepsilon}
\begin{document}
\title{Maxwell-Scalar device based on the electric dipole}
\author{D. Bazeia}%\email{bazeia@fisica.ufpb.br}
\affiliation{Departamento de F\'\i sica, Universidade Federal da Para\'\i ba, 58051-970 Jo\~ao Pessoa, PB, Brazil}
\author{M.A. Marques}%\email{marques@cbiotec.ufpb.br}
\affiliation{Departamento de Biotecnologia, Universidade Federal da Para\'\i ba, 58051-900 Jo\~ao Pessoa, PB, Brazil}
\affiliation{Departamento de F\'\i sica, Universidade Federal da Para\'\i ba, 58051-970 Jo\~ao Pessoa, PB, Brazil}
\author{R. Menezes}%\email{rmenezes@dce.ufpb.br}
\affiliation{Departamento de Ci\^encias Exatas, Universidade Federal
da Para\'{\i}ba, 58297-000 Rio Tinto, PB, Brazil}
\affiliation{Departamento de F\'\i sica, Universidade Federal da Para\'\i ba, 58051-970 Jo\~ao Pessoa, PB, Brazil}
\begin{abstract}
In this work we study the electric field of a dipole immersed in a medium with permittivity controlled by a real scalar field which is non-minimally coupled to the Maxwell field. We model the system with an interesting function, which allows the presence of exact solutions, describing the possibility of the permittivity to encapsulate the charges at very high values, giving rise to an effect that is not present in the standard situation. The results are of direct interest to applications for emission and absorption of radiation, and may motivate new studies concerning binary stars and black holes in gravity scenarios of current interest. 
\end{abstract}
%\date{\today}
\maketitle

160 years ago, Maxwell considered results described by Amp\`ere, Coulomb, Faraday and Gauss (and others) to write down a set of four equations that changed the physics in some important ways. These equations and the work of Hertz helped to set the basics for the generation and detection of electromagnetic waves. And more, with Lorentz and Einstein, they paved the way to develop the special theory of relativity.  Nowadays, electromagnetism is seen as an Abelian $U(1)$ gauge theory, described by the gauge field $A_\mu$ and the gauge-invariant tensor $F_{\mu\nu}=\partial_\mu A_\nu-\partial_\nu A_\mu$. Its Lagrange density is very well-known in physics, but in the present work we consider a much less known system, described by the addition of a real scalar field $\phi$ which is non-minimally coupled to the gauge field in the form 
\be\label{lmodel}
	\LL = \frac12\partial_\mu\phi\partial^\mu\phi - \frac{\veps(\phi)}{4}F_{\mu\nu}F^{\mu\nu} - A_\mu j^\mu,
\ee
in the presence of the (conserved) external current $j_\mu$. We notice that the Maxwell coupling to the scalar field is driven by the non-negative function $\veps(\phi)$, which governs the electric permittivity of the system. This was used before in \cite{lee,L} to describe a bag similar to the MIT \cite{mit} and SLAC \cite{slac} bag models. It was also considered in several distinct scenarios, in particular, in Refs. \cite{V1,V2,V3} to describe specific properties of vortices, in Refs. \cite{H1,H2,H22,H3} in connection with the AdS/CFT correspondence, to model insulators and metals within the holographic setup and in the context of the gauge/gravity duality, to investigate hydrodynamic behavior in a hot and dense strongly coupled quark-gluon plasma. And also in \cite{S1,S2,S3,S4}, to study dielectric Skyrme models and possible connections with the binding energies of nuclei and other related issues. 
In the present work, we go beyond the case of a single charge recently investigated in \cite{electric} and consider the presence of an electric dipole characterized by charge $e$ and distance $d=2a$.

In a way similar to \cite{V1,V2}, other works have investigated the Maxwell-Scalar coupling in the magnetic context in some recent investigations; see, e.g., Refs. \cite{M1,M2,M3,M4}, where the magnetic permeability is considered to induce internal structure to planar configurations of the vortex type. The above non-minimal coupling has also been considered in other scenarios, for instance, in Maxwell-Scalar models in the presence of nontrivial geometric backgrounds and also in Einstein-Maxwell-Scalar systems, with the scalar field giving rise to specific effects such as spontaneous scalarization, among others; see, e.g., \cite{G00,G0,G1,G12,G2,G3,G4} and references therein. Under specific coupling, it is of current interest to enlarge the system to construct black hole solutions with so-called spin-induced scalarization \cite{SI1,SI2}. 

In the present work, owing to applications in condensed matter, we suppose the system lives in the simplest case of a flat spacetime with $(2,1)$ dimensions. The spatial coordinates are described by $(x,y)$, and an electric dipole is symmetrically located in the $x$ axis, with the $y$ coordinate added to describe the plane of reference of the dipole. We start varying the action associated to the above Lagrange density to get the following equations of motion
\bes
\bal
\partial_\mu \partial^\mu \phi + \frac{1}{4}\veps_{\phi}F_{\mu\nu}F^{\mu\nu} =0, \\
\partial_\mu\left(\veps F^{\mu\nu}\right) = j^\nu,
\eal
\ees
where $\veps_{\phi} = d \veps/d\phi$. We then consider static configurations with time-independent $j^0$ and $j^i=0$, so there is no magnetic field involved in the system. The components of the electric field $\textbf{E}$ are defined as $E^i = (E_x,E_y) = F^{i0}$. In this case, the above equations of motion become
\bes
\bal\label{eomphi}
\nabla^2\phi + \frac12 \veps_\phi|\textbf{E}|^2=0, \\ \label{meqs}
\nabla\cdot (\veps\,\textbf{E})= j^0.
\eal
\ees
Eq. \eqref{meqs} is the Gauss' law in the presence of a generalized electric permittivity. Since we are dealing with static configurations, we can calculate a conserved energy density standardly, which has the form
\be\label{rho}
\rho = \frac12\left(\nabla\phi\right)^2 + \frac12\veps(\phi)|{\bf E}|^2 + A_0j^0.
\ee
The charge density of the dipole has the following form $j^0 = 2\pi e\left(\delta(x-a)\delta(y) - \delta(x+a)\delta(y)\right)$, and the electric field is obtained from Eq.~\eqref{meqs}, which leads us to the expression
\be\label{e}
\textbf{E} = \frac{e}{\veps(\phi)} \left(\frac{\hat{r}_+}{|\textbf{r}_+|}-\frac{\hat{r}_-}{|\textbf{r}_-|}\right),
\ee
where $\hat{r}_\pm$ identify the unit radial vectors in the direction that connects the charges $\pm e$ to the point of reference, given by $\hat{r}_\pm=\textbf{r}_\pm/|\textbf{r}_\pm|$, with
$\textbf{r}_{\pm} = \left(x\mp a\right) \hat{x} \,+ \, y\,\hat{y}$. We can see from Eq.~\eqref{e} that the electric field is modified by the function $\veps(\phi)$, which drives the electric permittivity of the medium in which the charges are placed. 

The presence of the dipole suggests that we change the Cartesian $(x,y)$ coordinates to the bipolar $(\xi,u)$ coordinates, in the form
$x = a\sinh\xi/(\cosh\xi-\cos u)$ and $ y=a\sin u/(\cosh\xi - \cos u),$
where $\xi\in\mathbb{R}$ and $u\in[-\pi,\pi)$ are now dimensionless coordinates. The line element is given by
\be\label{ds2}
ds^2 = dt^2-\frac{a^2}{\left(\cosh\xi-\cos u\right)^2}\left(d\xi^2+du^2\right),
\ee
and the electric field now reads
\be\label{ebip}
\textbf{E} = -\frac{e}{\veps(\phi)}\,\frac{\cosh\xi-\cos u}{a}\,\hat{\xi},
\ee
always pointing in the $\xi$ direction. In this expression, one notices that the $\cosh\xi$ in the numerator diverges as $\xi\to\pm\infty$. Nevertheless, the presence of the generalized electric permittivity $\veps(\phi)$ may be used to modify this behavior. In bipolar coordinates, the equation of motion of the scalar field \eqref{eomphi} becomes
\be\label{ephi}
\frac{\partial^2\phi}{\partial\xi^2} + \frac{\partial^2\phi}{\partial u^2}+ \frac{a^2}{2\left(\cosh\xi-\cos u\right)^2}\,\veps_\phi |\textbf{E}|^2=0.
\ee
The above Eqs.~\eqref{ebip} and \eqref{ephi} allow that we take $\phi=\phi(\xi)$, that is, the scalar field may not depend on $u$. In this case, for $\phi(\xi)$ one gets
\be\label{ephi1}
\frac{d^2\phi}{d\xi^2} = \frac{d}{d\phi}\left(\frac{e^2}{2\veps}\right).
\ee
Comparing this with the usual equation of motion for kinks in $(1,1)$ dimensions, which has the form $d^2\phi/dx^2 = dV/d\phi$ (see, e.g., Ref.~\cite{bazeia}), one can show that the electric permittivity may play the role of a potential. We define
\be\label{VP}
V(\phi) = \frac{e^2}{2\veps(\phi)},
\ee
and rewrite Eq. \eqref{ephi1} as
\be\label{eqkink}
\frac{d^2\phi}{d\xi^2} = \frac{dV}{d\phi}.
\ee

To bring to light novel information on the subject, let us now search for kinklike configuration for the scalar field. We then suppose the potential engenders two neighbour minima, say $\bar\phi_\pm$ such that $V(\bar\phi_\pm)=0$, which may be connected asymptotically by the field configuration $\phi(\xi)$. In this situation, since the potential and the permittivity are related by Eq.~\eqref{VP}, one can see that the permittivity diverges at the two zeroes of the potential, and this feature may be used to compensate the divergence caused by the $\cosh\xi$ in the electric field, as we have commented below Eq.~\eqref{ebip}. As usual, the above equation may be integrated to a first order one, in the form
\be\label{fokink}
\frac12\left(\frac{d\phi}{d\xi}\right)^2 = V(\phi).
\ee
To get the solution of the gauge field, $A_0$, we must take into account that $\textbf{E} = -\nabla A_0$, where the gradient is now given by
\be\label{nabla}
\nabla = \frac{\cosh\xi-\cos u}{a}\left(\hat{\xi}\,\frac{\partial}{\partial\xi} +\hat{u}\, \frac{\partial}{\partial u}\right).
\ee
Using the above expression in Eq.~\eqref{ebip} combined with Eq.~\eqref{VP}, we get that $\partial A_0/\partial u =0$ and $\partial A_0/\partial\xi = 2V/e$. So, we must have $A_0=A_0(\xi)$, which is given by
\be\label{a0phi}
\begin{split}
	A_0(\xi) &=\frac{2}{e}\int\,d\xi\, V(\phi(\xi))= \frac1e\int\,d\xi\left(\frac{d\phi}{d\xi}\right)^2.
\end{split}
\ee
Thus, by knowing the solution $\phi(\xi)$ obtained from Eq.~\eqref{fokink}, one can calculate $A_0(\xi)$ and then the electric field. The corresponding energy density is calculated from Eqs.~\eqref{rho}; it reads
\be
	\rho =\frac{\left(\cosh\xi\!-\!\cos u\right)^2}{2a^2}\left(\frac{d\phi}{d\xi}\right)^2\!\!+ \frac12\veps(\phi)|{\bf E}|^2 + A_0j^0,
\ee
The electric field may be written in terms of the potential \eqref{VP} as in Eq.~\eqref{ebip}, and this allows to write $\rho = \rho_{f} + \rho_{c}$, representing the contributions due to the scalar field and the charges, respectively. The two contributions are
\bes
\bal\label{rhof}
\rho_{f} &= \frac{\left(\cosh\xi-\cos u\right)^2}{a^2}\left(\frac12\left(\frac{d\phi}{d\xi}\right)^2 + V(\phi)\right),\\ \label{rhoc}
\rho_{c} &=2\pi eA_0\left(\delta(x-a)\delta(y) - \delta(x+a)\delta(y)\right).
\eal
\ees

To calculate the energy, we now have to use the equality $dx\,dy = a^2 du\,d\xi/(\cosh\xi-\cos u)^2$. From Eq.~\eqref{rhoc}, we get
\be\label{ec}\begin{aligned}
E_c&=2\pi e\left(A_0(x=a,y=0) - A_0(x=-a,y=0)\right)\\
&=2\pi e\left(A_0(\xi\to\infty) - A_0(\xi\to-\infty)\right).
\end{aligned}
\ee
The contribution due to the scalar field also depends on the specific form of $\phi(\xi)$. Here we dive further into the construction of stable structures attaining topological feature; we follow the Bogomol'nyi procedure \cite{bogo} and introduce another function, $W=W(\phi)$, such that 
\be\label{VW}
V(\phi) = \frac12\left(\frac{dW}{d\phi}\right)^2.
\ee
This allows that we rewrite $\rho_f$ in the form 
\be
\begin{aligned}
\rho_{f} = \frac{\left(\cosh\xi-\cos u\right)^2}{a^2}
	     \Bigg(\frac12\left(\frac{d\phi}{d\xi} \mp \frac{dW}{d\phi}\right)^2\! \pm \frac{dW}{d\xi}\Bigg),
\end{aligned}
\ee
such that 
\be
E_f=2\pi \int_{-\infty}^{\infty} d\xi \;\Bigg(\frac12\left(\frac{d\phi}{d\xi} \mp \frac{dW}{d\phi}\right)^2\! \pm \frac{dW}{d\xi}\Bigg).
\ee
This energy can then be minimized to the value
\be\label{eb}
2\pi |W(\phi(\xi\to\infty))-W(\phi(\xi\to-\infty))|,\\
\ee
when the field configuration $\phi(\xi)$ obeys one of the first order equations
\be\label{fo}
\frac{d\phi}{d\xi}= \pm \frac{dW}{d\phi}.
\ee
One can show that solutions of the above first order equations also solve the equation of motion \eqref{eqkink}. Moreover, we follow the lines of Refs.~\cite{prl,derrick,hobart} and make a rescale in the argument of the solution, in the form $\phi\to \phi_\kappa=\phi(\kappa\xi)$, to show that the solutions are stable under rescaling, so they evade Derrick's theorem. An extra important gain in the present case, is that we can write the function $A_0$ in Eq.~\eqref{a0phi} as 
\be\label{a0w}
A_0(\xi) = \pm\frac{1}{e}W(\phi(\xi)),
\ee
which makes the energy $E_c$ in Eq.~\eqref{ec} to be
\be\label{ecw}
E_c =2\pi \left|W(\phi(\xi\to\infty)) - W(\phi(\xi\to-\infty))\right|,
\ee
such that $E_c=E_f$. In Eq.~\eqref{a0w}, we have discarded a constant of integration because it plays no role in the energy as one can see from the above expression. In the case of a single charge that we investigated in Ref.~\cite{electric}, the constant of integration can be seen as a reference for the energy. Here, since we have two charges of opposite signs, it has no importance and can be set to zero. The procedure naturally reveals how to remove the singular behavior of the energy associated to the standard case ($\veps=1$) due to the presence of the two charges. For simplicity, we take $e=1$ from now on.

%%%%%%%%%%%%%%%%
\begin{figure}[t!]
\centering
\includegraphics[width=2.8cm]{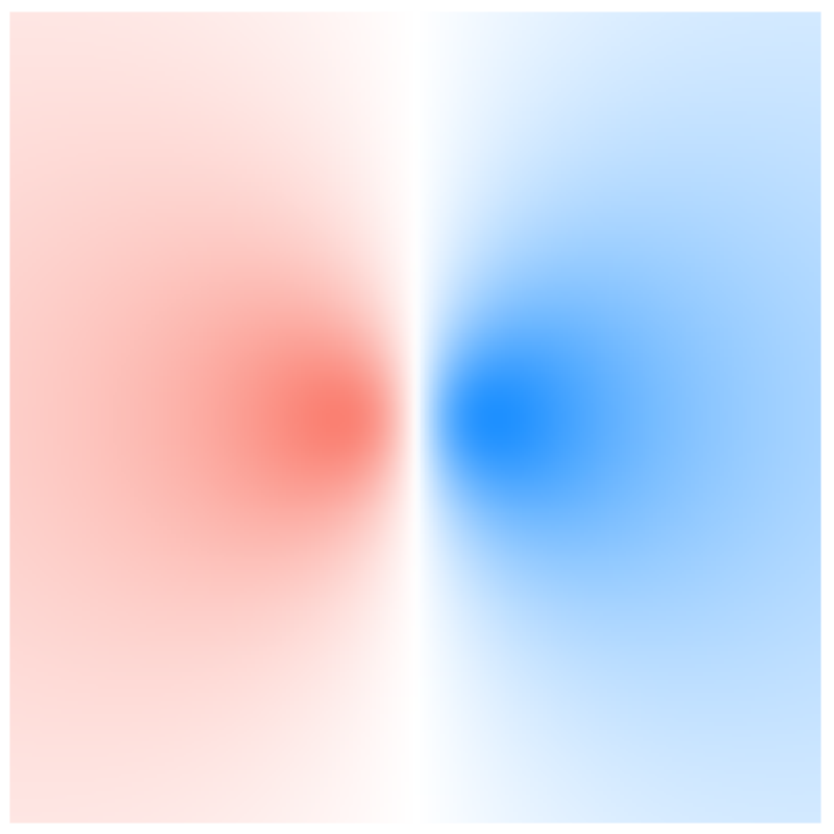}
\includegraphics[width=2.8cm]{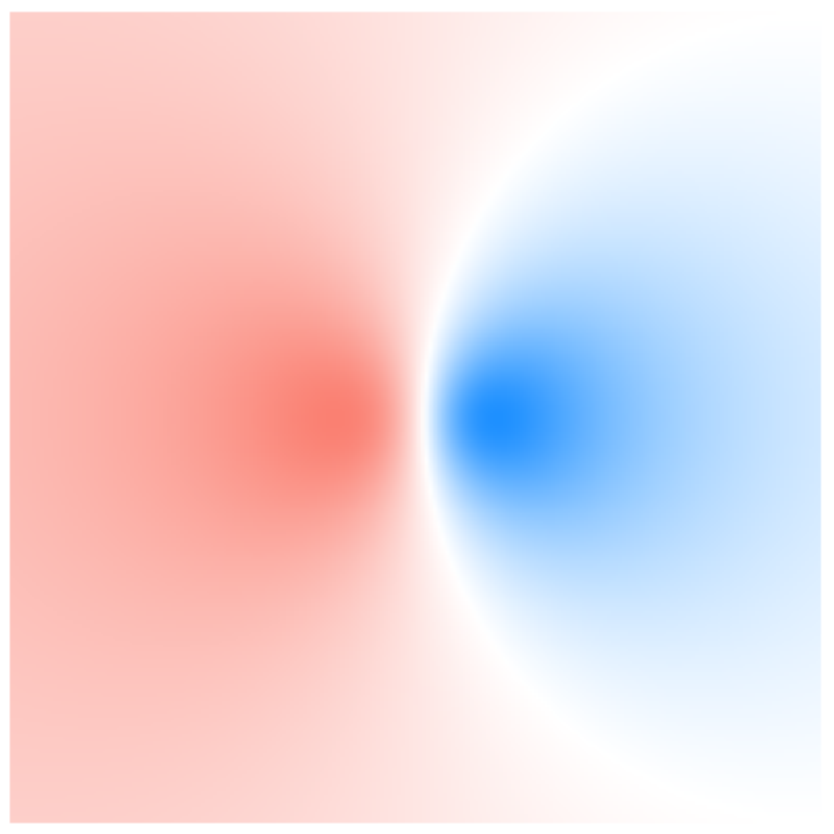}
\includegraphics[width=2.8cm]{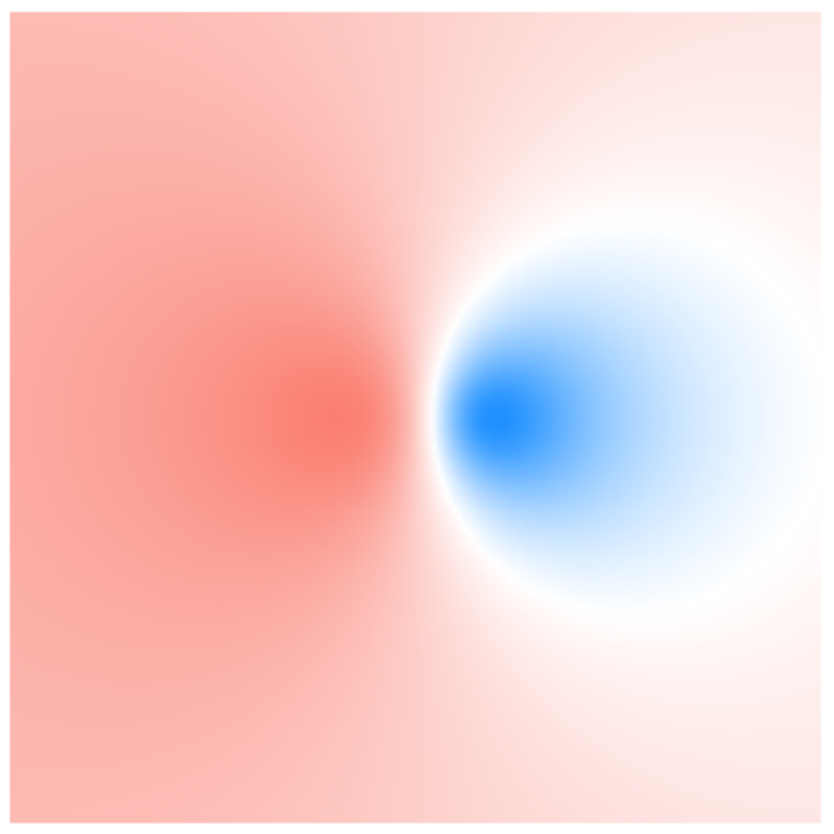}\\
\vspace{0.3cm}
\includegraphics[width=2.8cm]{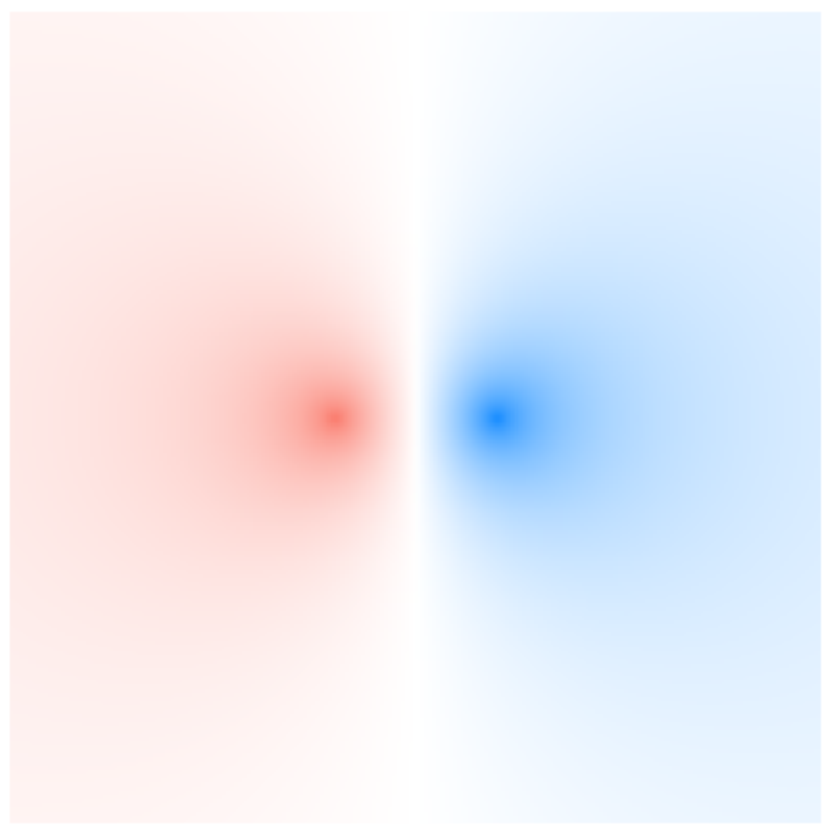}
\includegraphics[width=2.8cm]{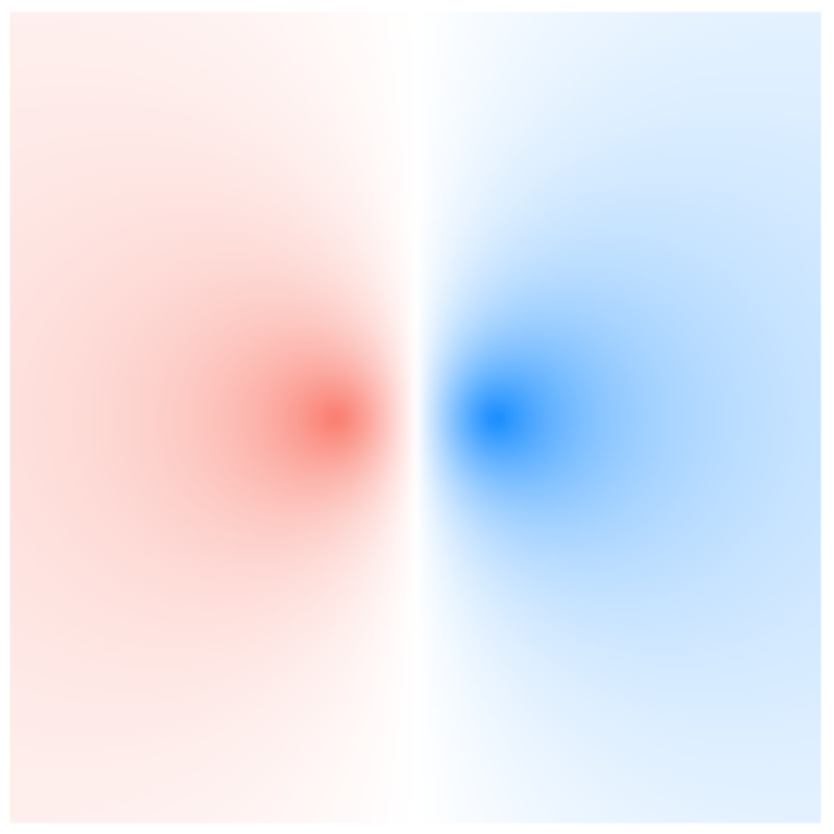}
\includegraphics[width=2.8cm]{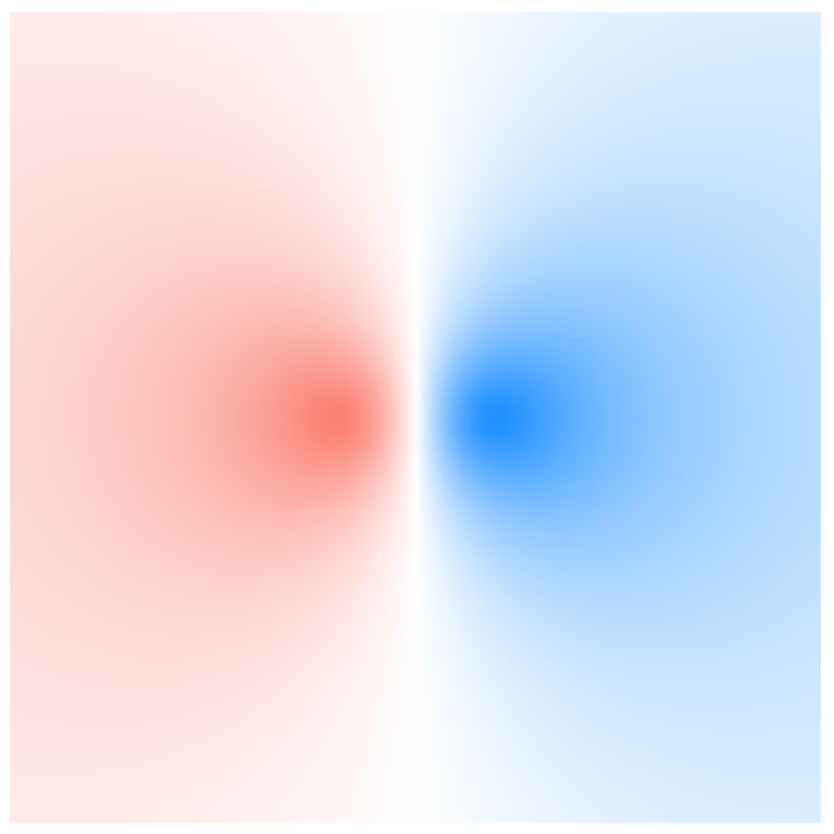}
\caption{The solution in Eq.~\eqref{kinkbipolar} is depicted in the plane $(x,y)$ for $a=1$ and $v=1$. In the top panels we take $\lambda=1$ and $\xi_0=0$, $0.2$, and $0.4$, from left to right. In the bottom panels we take $\xi_0=0$ and $\lambda=0.4$, $0.6$ and $0.8$, also from left to right. The red color stands for $\phi=-1$, the white color represents $\phi=0$ and the blue denotes $\phi=1$.}
\label{fig1}
\end{figure}
%%%%%%%%%%%%%%%%

To illustrate the above procedure, we consider the model described by the polynomial function
\be\label{modelW}
W(\phi)=\lambda v \phi-\frac{\lambda}{3v}\phi^3,
\ee
where $\lambda$ and $v$ are positive parameters. In this case, the potential in Eq.~\eqref{VW} and the electric permittivity in Eq.~\eqref{VP} change to
\bal\label{v1}
  V(\phi) &= \frac{\lambda^2v^2}{2}\!\left(1-\frac{\phi^2}{v^2}\right)^2,\\
  \label{perm1}
 \veps(\phi)&=\left(\lambda^2v^2\left(1-\frac{\phi^2}{v^2}\right)^2\right)^{-1}.
\eal
The above $V(\phi)$ has minima at $\bar\phi_\pm=\pm v$ and a local maximum at $\phi=0$, such that $V(0)=\lambda^2v^2/2$. Also, the permittivity is constant at $\phi=0$ ($\varepsilon(0)=1/\lambda^2 v^2$) and increases to higher and higher values as the scalar field approaches the minima $\pm v$. The presence of the double-well potential is mandatory to provide the kinklike configuration to be used to regularize the behavior of the electric field at the two charges. The model is also motivated by the Landau-Ginzburg-Devonshire theory for ferroelectric materials, which may describe negative capacitance and improve the energy efficiency of conventional electronics (see, e.g., \cite{nature} and references therein). The first order equation \eqref{fokink} with positive sign now reads
\be
\frac{d\phi}{d\xi} = \lambda v\!\left(1-\frac{\phi^2}{v^2}\right),
\ee
and the solution which connects the aforementioned minima of $V(\phi)$ is
\be\label{kinkbipolar}
\phi(\xi) = v\tanh\!\left(\lambda (\xi-\xi_0)\right),
\ee
where $\xi_0$ determines the point in the $\xi$ axis where the solution changes its sign. The situation is similar to the case of kinks in the real line \cite{bazeia}, and $\xi_0$ is an integration constant which does not modify the energy of the solution. The behavior of this analytical result is depicted in Fig.~\ref{fig1} in the $(x,y)$ plane for $a=1$ and $v=1$, and for several values of $\xi_0$ and $\lambda$. In this figure, the white region represents $\phi=0$, which requires $\xi=\xi_0$. In the top panel the circle appears in accordance with $(x-a\coth\xi_0)^2 + y^2 = a^2/\sinh^2\xi_0$, so it has radius $a/\sinh\xi_0$ and is centered at the point $(\tilde {x},\tilde{y})=(a\coth\xi_0,0)$. In the present context, $\xi_0$ can be used to induce an asymmetry in the system, which also appears in the permittivity and in the electric field, as we show below. The asymmetry can be used for applications of practical interest.

Moreover, the contribution of the field for the energy density $\rho_f$ is given by Eq.~\eqref{rhof}. We have
\be\label{rhobipolar}
\rho_f =  \frac{\left(\cosh\xi-\cos u\right)^2}{a^2}\,\lambda^2v^2\sech^4(\lambda (\xi-\xi_0)).
\ee
We can use Eqs.~\eqref{eb} and \eqref{ecw} to show that $E_f=E_c=8\pi\lambda v^2/3$, giving the total energy $E=16\pi\lambda v^2/3$.

The electric potential and the electric field are obtained from Eqs.~\eqref{a0w} and \eqref{ebip}. The electric potential in Eq.~\eqref{a0w} becomes
\be
A_0(\xi)=\lambda v^2\!\!\left(\!\tanh(\lambda(\xi\!-\!\xi_0))\! -\!\frac13\tanh^3(\lambda(\xi\!-\!\xi_0))\!\right)\!,
\ee
and has a smooth behavior as one approaches the electric charges: $A_0(\xi\to\pm\infty)\to\pm2\lambda v^2/3$. The electric field, however, has the form
\bal\label{electric}
\textbf{E} &= -\frac{\lambda^2v^2}{a}(\cosh\xi-\cos u)\,\sech^4(\lambda(\xi\!-\!\xi_0))\,\hat{\xi}.
\eal
The presence of the generalized electric permittivity in Eq.~\eqref{perm1}, may now contribute to modify the singular character of the standard Coulomb's law at the location of the charges. To see this, one notices that near the charges one has $\xi_\pm\approx\pm\ln\left(2a/\sqrt{(x\mp a)^2+y^2}\right)$, so we can expand the above expression of the electric field to get
\be
|\textbf{E}|\approx \frac{16^{(1-\lambda)}\lambda^2v^2}{a^{4\lambda}}\left((x\mp a)^2+y^2\right)^{2\lambda-1/2}.
\ee
The quantity $2\lambda-1/2$ appears from the interplay between the bipolar coordinate system and the field configuration that solves the model.

We then see that the behavior of the electric field depends on $a$, $v$ and $\lambda$, with $\lambda_c=1/4$ being a critical value, for which the electric field is ill-defined at the location of the charges due to the presence of the unit vector $\hat{\xi}$. At the charges, for $\lambda<\lambda_c$ the electric field is divergent, as it is in the standard case, and for $\lambda>\lambda_c$ the electric field vanishes. Since the case of $\lambda$ greater than $1/4$ induces a novel scenario, we then focus on $\lambda> 1/4$ from now on. 

%%%%%%%%%%%%%%%%
\begin{figure}[t!]
\centering
\includegraphics[width=2.8cm]{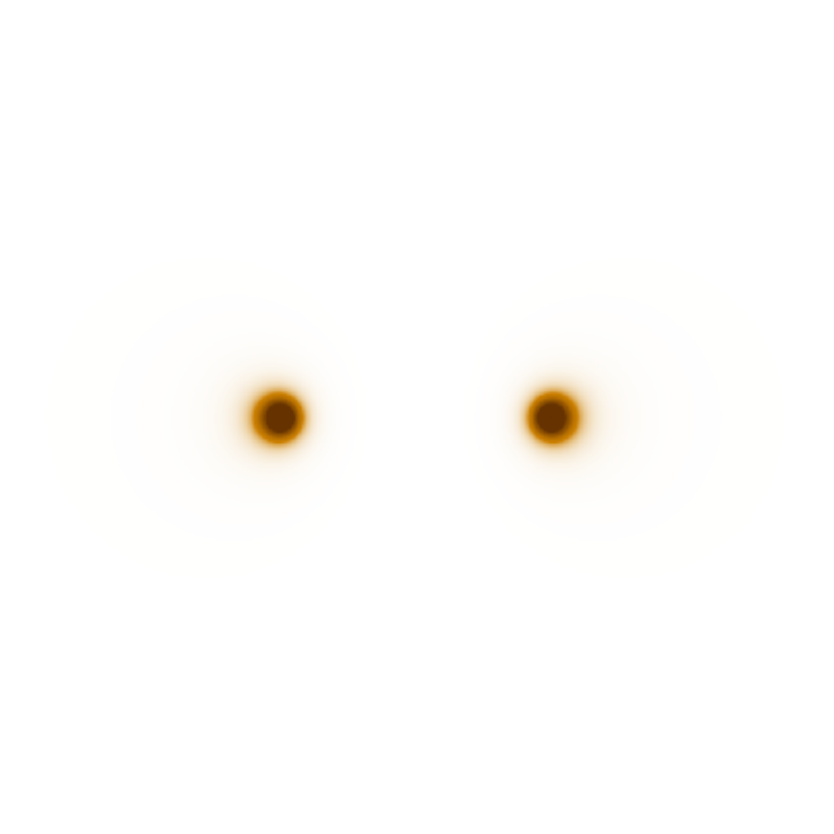}
\includegraphics[width=2.8cm]{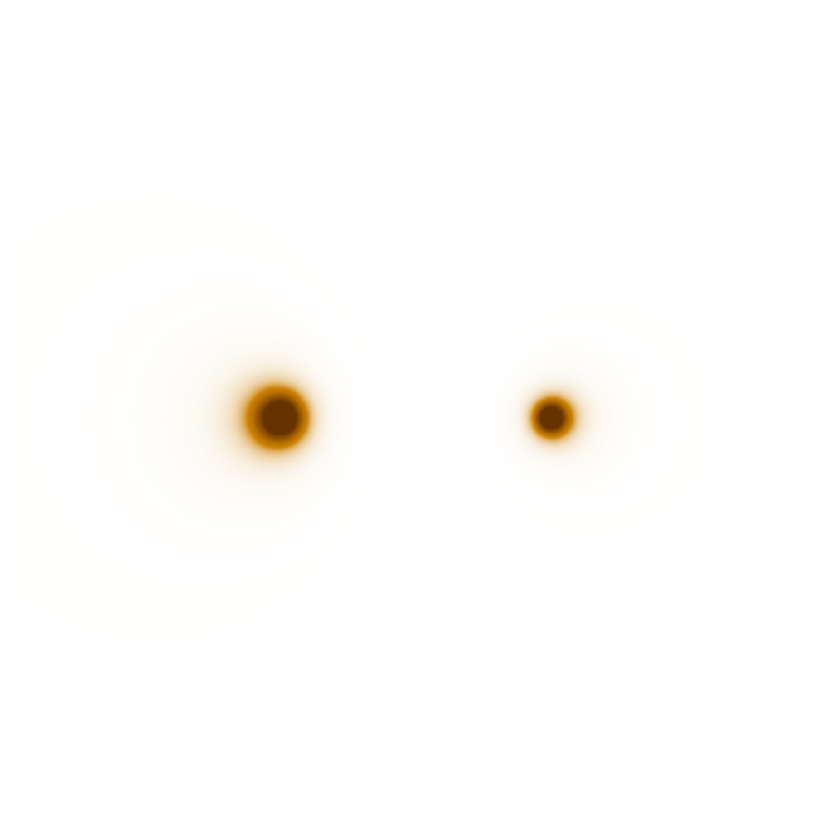}
\includegraphics[width=2.8cm]{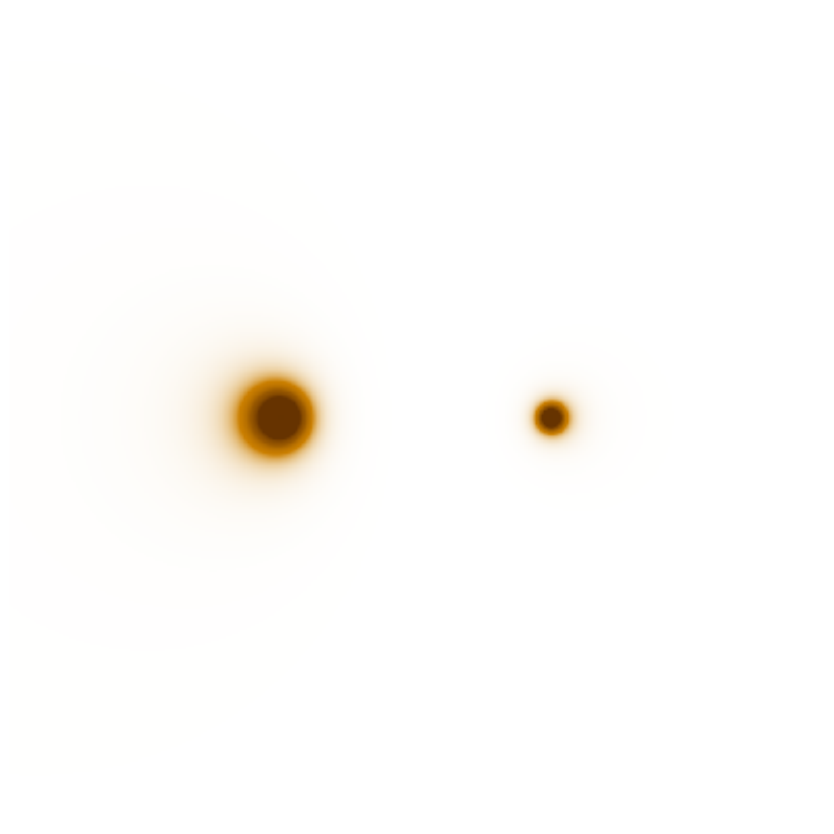}
\caption{The electric permittivity of Eq. \eqref{per}, depicted in the $(x,y)$ plane for $a=1$, $v=1$, $\lambda=1$ and for $\xi_0=0,0.2$ and $0.4$, from left to right. Here, the white color represents the vacuum $(\varepsilon=1)$, and the darkness of the brown color increases with the enlargement of the permittivity.}
\label{fig3a}
\end{figure}
%%%%%%%%%%%%%%%%
%%%%%%%%%%%%%%%%%%%%%%%%
\begin{figure}[t!]
\centering
\includegraphics[width=2.8cm]{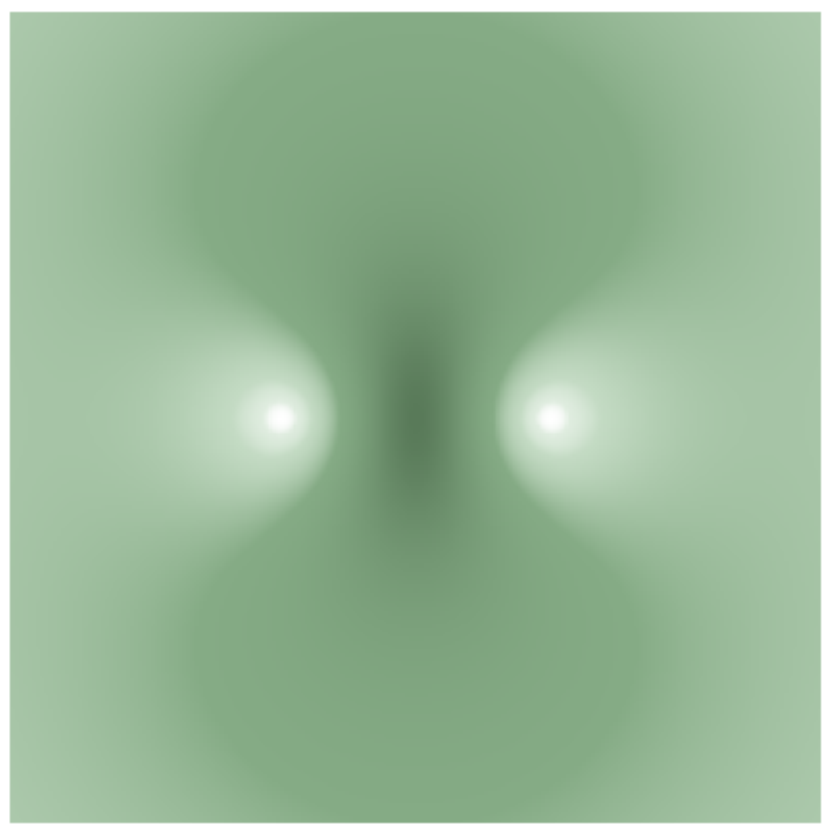}
\includegraphics[width=2.8cm]{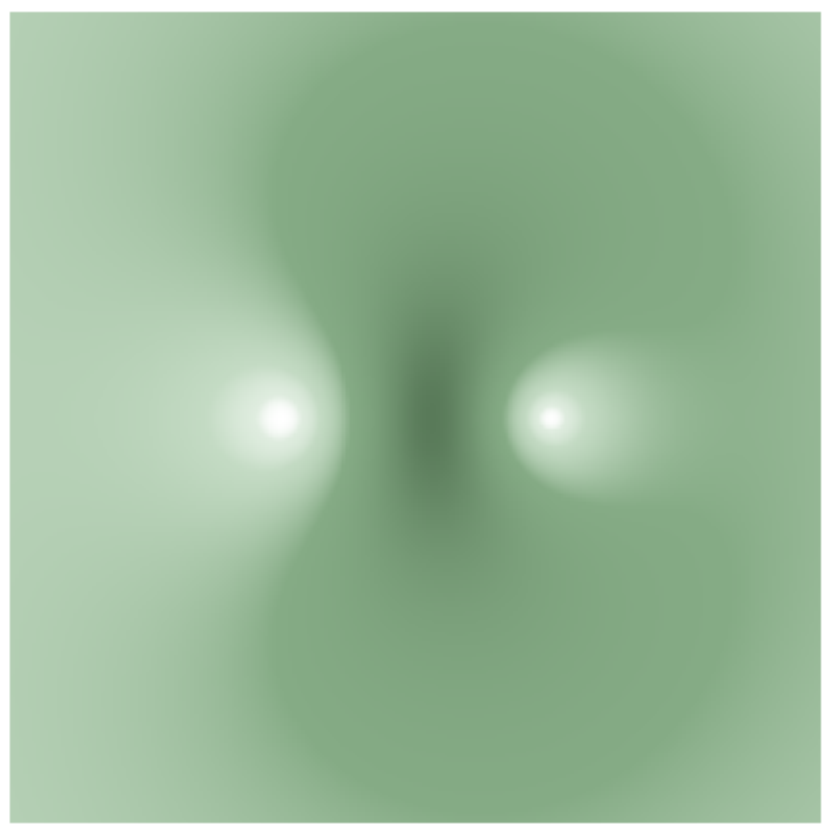}
\includegraphics[width=2.8cm]{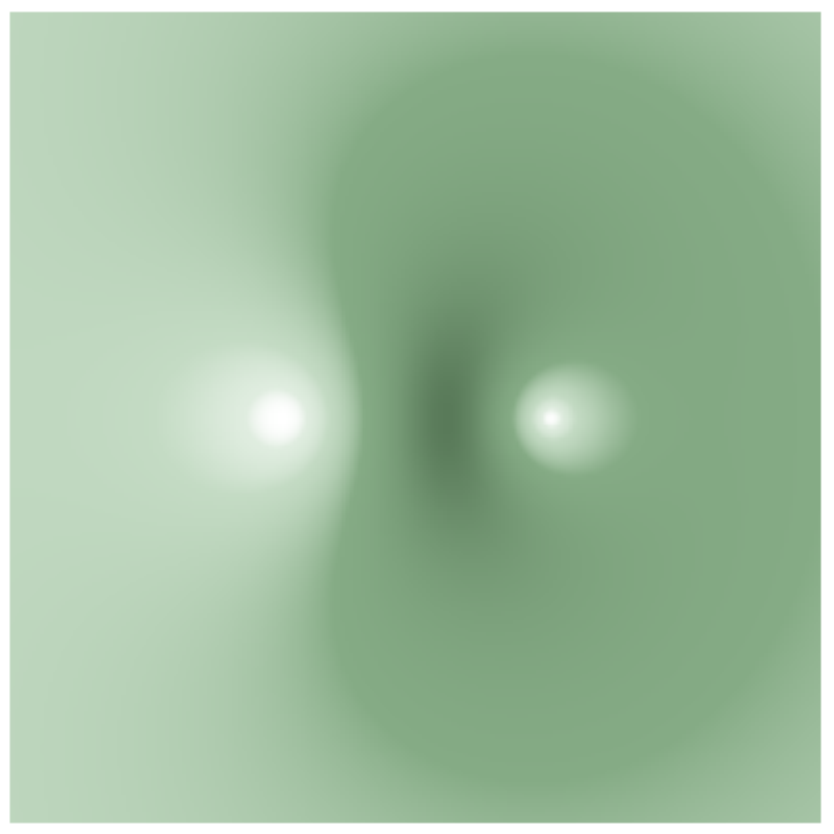}\\
\vspace{0.3cm}
\includegraphics[width=2.8cm]{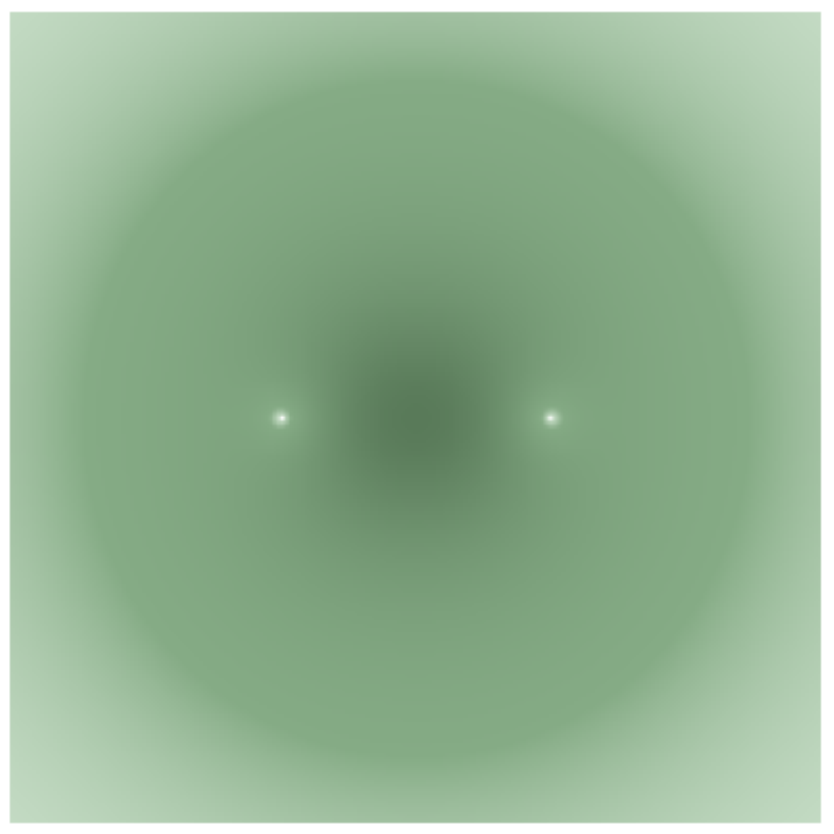}
\includegraphics[width=2.8cm]{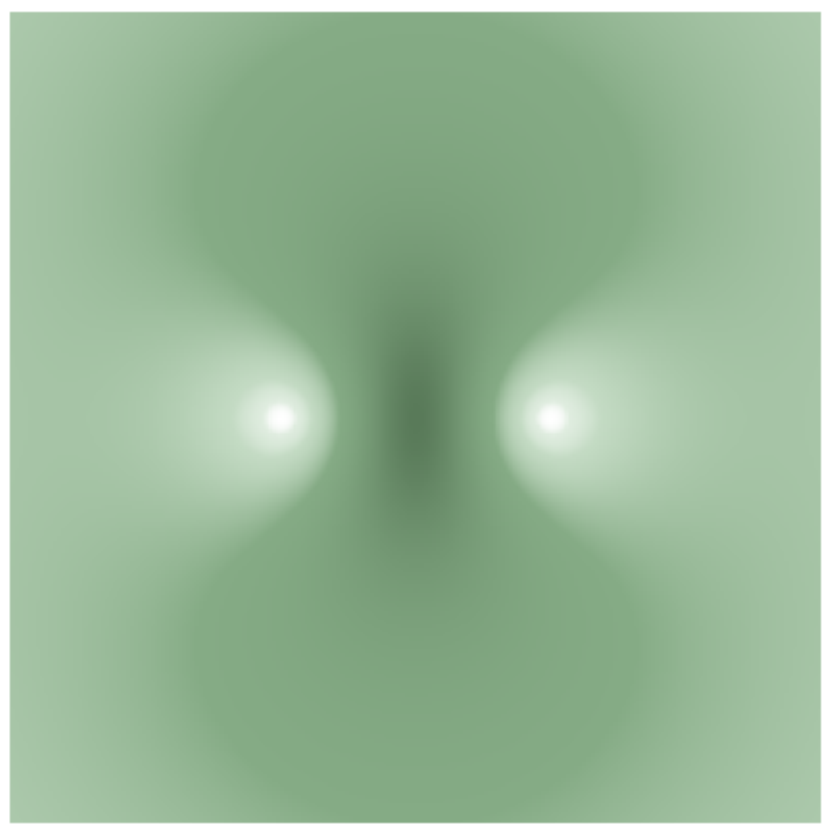}
\includegraphics[width=2.8cm]{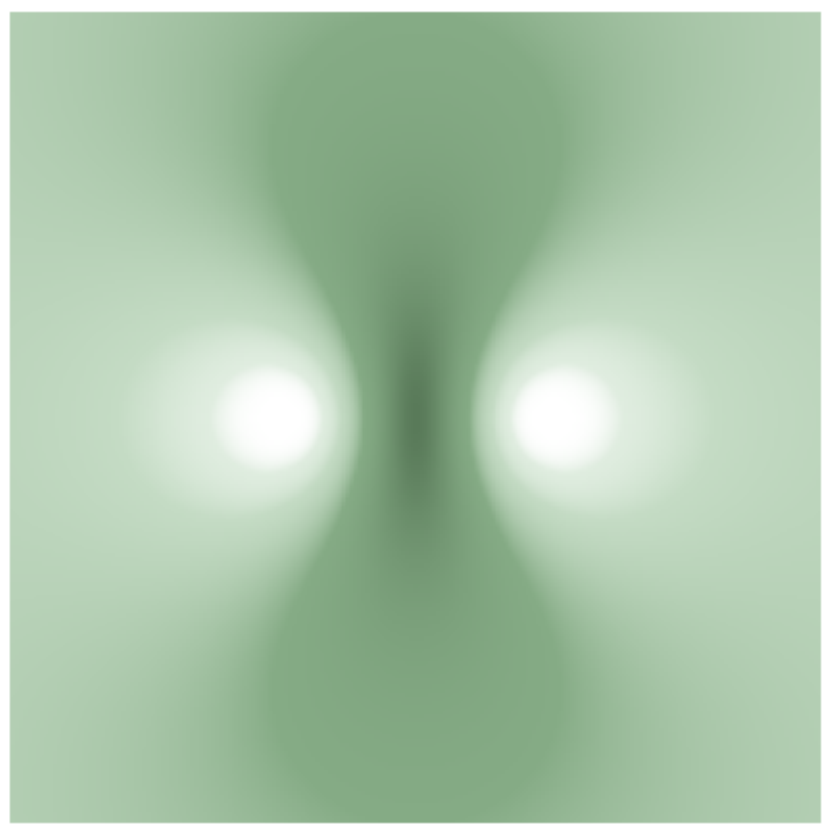}
\caption{The electric field in Eq.~\eqref{electric} is depicted in the $(x,y)$ plane for $a=1$ and $v=1$. In the top panels, we take $\lambda=1$ and $\xi_0=0$, $0.2$, and $0.4$, from left to right. In the bottom panels, we take $\xi_0=0$ and $\lambda=1/2$, $1$ and $3/2$, also from left to right. The color white represents vanishing electric field, and the shades of green are used to describe its variation, with the darkest standing for its maximum.}
\label{fig3}
\end{figure}
%%%%%%%%%%%%%%%%%%%%%%%

In the above model, for the field configuration shown in Eq. \eqref{kinkbipolar}, the electric permittivity becomes
\be\label{per}
\veps(\xi)=\lambda^{-2}v^{-2}\cosh^4(\lambda(\xi-\xi_0)).
\ee
It also introduces an asymmetry for $\xi_0\neq0$, so in Fig. \ref{fig3a} we depict this permittivity using $a=1$, $v=1$ and $\lambda=1$, for some values of $\xi_0$, remembering that $\varepsilon=1$ represents the permittivity of the vacuum, which in the figure is identified with the color white. To illustrate the behavior of the electric field, in Fig. {\ref{fig3}} we depict the analytical results in Eq.~\eqref{electric} in the $(x,y)$ plane for $a=1$, $v=1$, and for several values of $\xi_0$ and $\lambda$.

In summary, in this work we have investigated a model in which a single real scalar field is coupled to the Maxwell field through an electric permittivity. We studied the case in which two point charges with opposite signs are symmetrically fixed in the $x$ axis, representing an electric dipole, using an interesting model of permittivity, inspired by the Landau-Ginzburg-Devonshire theory for ferroelectric materials \cite{nature}, which allows for the presence of exact solutions. The study unveils an interesting behavior of the electric field, showing that it may vanish at the positions of the electric charges. This is very different from the standard situation, and is attained when the two charges of the electric dipole are encapsulated inside regions with permittivity that increases to larger and larger values, as we approach the charges. The results suggest an interesting modification of the standard electric dipole, which we believe is not hard to construct in the laboratory, if one uses ceramic or other material with very high dielectric constants; see, e.g., Refs. \cite{VH1,VH2,VH3,VH4,VH5} for more details on the use of ceramic \cite{VH1,VH2}, dielectric film \cite{VH3,VH4} and gel \cite{VH5} elements. The encapsulated dipole engenders new physics, so it may be used for practical purposes of current interest in applications including emission and absorption of radiation. Moreover, as one can see from Refs. \cite{electric,G4}, in the Maxwell-Scalar model we can trade off the electric charge for a nontrivial gravitational background. In this sense, the above results are of direct interest to study binary stars, binary black holes and other similar astrophysical systems. In particular, it seems of current interest to investigate how the process of scalarization works for binary black holes systems \cite{G0,G1,SI1,SI2,mer1,mer2}. 

The Maxwell-Scalar system considered in \eqref{lmodel} may also be studied to unveil other possibilities, for instance, the modification of the model considered in Eq. \eqref{modelW} to describe different features of the field configuration that solve the equation of motion, inducing novel possibilities of applications concerning negative capacitance effects in ferroelectric materials \cite{nature,new}. These and other related issues are presently under consideration, and we hope to report on them in the near future.

\acknowledgements{This work is supported by the Brazilian agencies Conselho Nacional de Desenvolvimento Cient\'ifico e Tecnol\'ogico (CNPq), grants Nos. 404913/2018-0 (DB), 303469/2019-6 (DB) and 306504/2018-9 (RM), Paraiba State Research Foundation (FAPESQ-PB) grants Nos. 0003/2019 (RM) and 0015/2019 (DB and MAM) and by Federal University of Para\'iba (PROPESQ/PRPG/UFPB) project code PII13363-2020.}
%%%%%%%%%%%%%%%%%%%%%%%%%%%%%%%%%

\end{document}